# Numerical model of solid phase transformations governed by nucleation and growth. Microstructure development during isothermal crystallization.

Jordi Farjas and Pere Roura

GRMT, Department of Physics, University of Girona, Campus Montilivi, Edif. PII, E17071 Girona, Catalonia, Spain

A simple numerical model which calculates the kinetics of crystallization involving randomly distributed nucleation and isotropic growth is presented. The model can be applied to different thermal histories and no restrictions are imposed on the time and the temperature dependencies of the nucleation and growth rates. We also develop an algorithm which evaluates the corresponding emerging grain size distribution. The algorithm is easy to implement and particularly flexible making it possible to simulate several experimental conditions. Its simplicity and minimal computer requirements allow high accuracy for two- and three-dimensional growth simulations. The algorithm is applied to explore the grain morphology development during isothermal treatments for several nucleation regimes. In particular, thermal nucleation, pre-existing nuclei and the combination of both nucleation mechanisms are analyzed. For the first two cases, the universal grain size distribution is obtained. The high accuracy of the model is stated from its comparison to analytical predictions. Finally, the validity of the Kolmogorov-Johnson-Mehl-Avrami model is verified for all the cases studied.

Corresponding author: jordi.farjas@udg.es

# I. INTRODUCTION

Transformed volume and grain morphology development due to solid-phase crystallization depend on two kinetic parameters: growth and nucleation rates. These parameters can be obtained from the morphological evolution observed by microscopy.[1] In addition, thermoanalytical techniques provide a simple and rapid way to measure the crystallization kinetics.[2] The aim of these kinds of analyses is to predict the crystallization behavior in order to define thermal treatments suitable to achieve a particular microstructure.

Crystallization of amorphous materials and other solid-phase transformation are generally described by the Kolmogorov-Johnson-Mehl-Avrami (KJMA) theory.[3-9] The KJMA theory is based on the assumption of spatially random nucleation and isotropic growth. Under these assumptions Avrami demonstrated that:[4]

$$\frac{d\alpha(t)}{1-\alpha(t)} = d\alpha_{ex}(t) \tag{1}$$

where $\alpha(t)$ is the transformed fraction at a time $t$ and $\alpha_{ex}(t)$ is the extended transformed fraction, i.e. the resulting transformed fraction if grains grow through each other and overlap without mutual interference:

$$\alpha_{ex}(t) = \int_0^t I(\tau) v_{ex}(\tau,t) d\tau . \tag{2}$$

In Eq. (2), $I$ is the nucleation rate per unit volume and $v_{ex}(\tau,t)$ is the extended volume at time $t$ of a single nucleus formed at time $\tau$

$$v_{ex}(\tau,t) = \sigma \left( \int_\tau^t G(z) dz \right)^m \tag{3}$$

where $\sigma$ is a shape factor (e.g. $\sigma = 4\pi/3$ for spherical grains), $G$ is the growth rate and $m$ depends on the growth mechanism[2] (e.g. $m=3$ for 3D growth). The integration of Eq. (1) gives:

$$\alpha(t) = 1 - \exp[-\alpha_{ex}(t)] \tag{4}$$

Although some authors[10] have cast doubts on the correctness of the KJMA theory, the relationship between $\alpha(t)$ and $\alpha_{ext}(t)$ of Eq. (4) is exact.[11] Recent numerical simulations[8,9,12-15] have confirmed it for several particular cases (a noteworthy analysis is given in ref. 12). The KJMA theory also holds in case of anisotropic growth provided that the grains have a convex shape and are aligned in parallel.[16] Moreover, the KJMA theory provides a good approximation when the anisotropy is moderate[13] or for soft-impingement and nonrandom nucleation.[17]

Unfortunately, as far as we know, analytically exact solutions for the transformed fraction, $\alpha(t)$, are restricted to three particular situations under isothermal conditions: time-independent growth and nucleation rates, time-independent growth rate and nucleation rate proportional to a power of time,[18] and preexisting nuclei (site saturation). Recently, a quasi-exact solution of the KJMA model has been obtained under continuous heating conditions.[19] In contrast, many transformations are governed by time-dependent nucleation and growth rates and non-isothermal heat treatments are a common practice. Thus, numerical calculations are needed to simulate these general cases. The main difficulty in numerically solving Eq. (2) is the dependence on the time history through $\tau$. A common solution is the analytical development and numerical integration of Eq. (2) for a particular set of conditions. Conversely, the number of general numerical solutions is quite sparse. Yinnon et al.[20] developed a simple method under the assumption of linear cooling or heating rate. Besides, Krüger[21] followed a different approach for non-isothermal transformations. The validity of the latter numerical solution is limited to some particular cases, as will be commented on in this paper.

The particular kinetic conditions of a phase transformation have an essential effect on the emerging grain morphology and, therefore, the material properties. Thus, several computer simulations have been developed to predict the resulting microstructure.[22] These simulations can be classified into two main groups: those based on a Monte Carlo method[23,24] and those based on cellular automata.[25,26] A common drawback of both approaches is the accumulative error at each evaluation step related to the spatial resolution resulting from space discretization. In general, the spatial resolution should be chosen high enough to reduce this error and the simulated volume should be high enough to reduce the statistical error related to the finite number of nuclei.[13] The problem is that a high spatial resolution limits the space extent. The problems related to the finite extent can be diminished by using periodic boundary conditions and by performing several runs to minimize the statistical error. Consequently, simulations require a significant amount of CPU time and memory.[14] Thus, three dimensional (3D) growth simulations are scarce and require the use of high performance computers. Furthermore, since a limited number of nuclei have a pronounced effect on the accuracy of grain size distribution calculation, at present, the calculation of accurate grain-size distribution continues to be an open problem. Pineda et al.[14,27] developed a method which allows the calculation of the grain size distribution in the framework of the KJMA theory and assuming a time dependent mean growth rate. Although their approach can be applied to a large number of cases, it does not provide an image of the final microstructure and the accuracy of the predicted grain-size distributions has been tested only indirectly.[14]

In this paper we introduce a new approach to address both problems: *a)* the calculation of $\alpha(t)$ under the assumptions of the KJMA theory and *b)* the calculation of the microstructure. First, we present a very simple numerical method that obtains $\alpha(t)$ for any particular case. The method is based on the calculation of $\alpha_{ext}(t)$ from a discrete set of nuclei. Once the extended volume is known, Eq. (4) calculates the transformed fraction. Thanks to its simplicity and flexibility, the numerical solution can be used to extract kinetic data from experiments. The validity of the model is tested for the case of continuous nucleation and growth under isothermal and continuous heating conditions. In addition, the model is applied to the analysis of the effect of partial crystallization prior to isothermal treatments.

Afterwards, we introduce an algorithm that calculates the microstructure and grain-size distribution. The algorithm computes the microstructure from the previous numerical calculation of the nuclei extended volume. In contrast with the previous calculation, the actual transformed fraction is not obtained from Eq. (4). Indeed, with this algorithm, the microstructure and transformed fraction are calculated directly from nucleation, growth and impingement of the individual grains. Thus, its applicability is not restricted by the KJMA assumptions. Since our approach reaches high accuracy in short computation times with minimal computer requirements, the microstructure for 3D growth can be easily calculated. Indeed, we will compute the evolution of the transformed fraction and the grain-size distribution under isothermal conditions and for several nucleation mechanisms. The accuracy of our approach is tested against some analytical results. In particular, the prediction of the mean grain size is excellent and indicates that the grain-size distributions obtained in this paper are very accurate. Finally, the correctness of the KJMA theory is verified in all the cases.

## II. NUMERICAL CALCULATION OF THE KJMA KINETIC EQUATIONS

Our numerical approach follows Kolmogorov's[7] development; for a finite volume $V$ of the parent phase, the extended volume is:

$$\alpha_{ex}(t) = \sum_{i=1}^{N(t)} \frac{v_{ex,i}(t)}{V} \qquad (5)$$

The previous summation covers all the grains. Since the volume of the parent phase is finite, the number of grains $N(t)$ is also finite. Thus, the actual transformed fraction can be derived from Eq. (4) provided that the volume of any arbitrary grain is much smaller than $V$. The latter condition is fulfilled if the number of grains is large enough. (According to appendix C the number of grains is equal to the ratio between $V$ and the mean grain volume)

The numerical calculation consists in the creation of an array which, for each single nucleus, $i$, stores its radius $r_i$. To avoid an accumulative rounding error related to the calculation of $N(t_j)$, at each time step, $j$, first we calculate the total number of grains, $N(t_j)$, and then the number of nuclei created, $\Delta N(t_j)$:

$$N(t_j) = V \cdot \int_0^{t_j} I(z)dz, \quad \Delta N(t_j) = N(t_j) - N(t_{j-1}) \qquad (6)$$

then, for all the previous created grains, $i$, their radius, $r_i(t_j)$, are updated:

$$r_i(t_j) = r_i(t_{j-1}) + \Delta r(t_j), \quad i = 1,...,N(t_{j-1}) \qquad (7)$$

where $\Delta r(t_j)$ is the radius growth in the time interval $t_j - t_{j-1}$:

$$\Delta r(t_j) = \int_{t_{j-1}}^{t_j} G(z)dz \qquad (8)$$

Note that $r$ is the radius of the extended volume of a nucleus, i.e., the radius supposing that the grain grows free.

For the grains created during the time interval $t_j - t_{j-1}$ an average radius is assumed:

$$r_i(t_j) = \frac{\Delta r(t_j)}{2} + r_0, \quad i = N(t_{j-1})+1,...,N(t_j) \qquad (9)$$

where $r_0$ is the critical germ size. In our calculations we will assume that $r_0$ is negligible. A more accurate calculation is obtained if the actual radius is calculated for each new nucleus:

$$r_i(t_j) = \int_{\tau_i}^{t_j} G(z)dz + r_0, \quad i = N(t_{j-1})+1,...,N(t_j) \qquad (10)$$

where $\tau_i$ is the creation time for the nucleus *i*. Besides, it is not necessary for the time interval to be constant; on the contrary, a more efficient simulation is obtained when the time interval is chosen such that a constant growth, $\Delta r$, is imposed.

$$\int_{t_{j-1}}^{t_j} G(z)dz = \Delta r \tag{11}$$

Finally, the extended fraction can be obtained according to Eq. (5):

$$\alpha_{ex}(t_j) = \frac{1}{V} \sum_{i=1}^{N(t_j)} \sigma[r_i(t_j)]^m \tag{12}$$

The simplicity of our approach relies on the fact that we solve the time history dependence by storing the radius for each nucleus. Moreover, no assumptions have been made on the time dependence of both *I* and *G*, so the numerical solution is general.

## III. ACCURACY OF THE NUMERICAL CALCULATION

To check the accuracy of our numerical calculation, test runs were done for 3D growth during isothermal and continuous heating, for which there exist an exact[7] and a quasi-exact solution,[19] respectively. The calculations have been done for the particular *G* and *I* values of amorphous silicon:

$$I = I_0 \exp(-E_N / k_B T) \text{ and } G = G_0 \exp(-E_G / k_B T) \tag{13}$$

where *T* is the temperature in Kelvin and $k_B$ is the Boltzmann constant. In Table I we summarize the parameters found in the literature.[1]

When nucleation and growth rates follow an Arrhenius dependence on temperature, the exact solution for the isothermal regime is:

$$\alpha = 1 - \exp[-k^{m+1}(t - t_0)^{m+1}], \quad k = \left(\frac{\sigma I G^m}{m+1}\right)^{1/{m+1}} = k_0 \exp\left(-\frac{E}{k_B T}\right) \tag{14}$$

where $E \equiv \dfrac{E_N + mE_G}{m+1}$ is an average activation energy and $t_0$ is the initial time. On the other hand, the quasi-exact solution for the continuous heating case is:[19]

$$\alpha = 1 - Exp\left\{-\left[k_0 C \frac{E}{\beta k_B} p\left(\frac{E}{k_B T}\right)\right]^{m+1}\right\} \tag{15}$$

where $p(x) \equiv \int_x^\infty \dfrac{\exp(-u)}{u^2} du$ and $C \equiv \left(\dfrac{(m+1)! E^{m+1}}{\prod_{i=0}^m (E_N + i E_G)}\right)^{1/{m+1}}$.

For the whole range of $\alpha$ and when the total number of nuclei is of the order of $10^5$, the deviations from the exact solutions are smaller than $10^{-5}$ for the isothermal case, while for the non-isothermal case the discrepancies are smaller than 0.01. Actually, the larger discrepancy for the non-isothermal case is due to the analytical solution and not to the present calculation. In fact, simulations performed when the quasi-exact solution becomes exact (taking the same activation energy for growth and nucleation[19]) resulted in deviations smaller that $10^{-6}$. In addition, similar minor discrepancies were also obtained for the case where all nuclei appear at $t = t_0$ - the site saturation case.[2]

Often a thermal treatment consists of an initial constant heating period followed by an isothermal step. A widespread approximation to this problem consists in introducing a virtual initial time $t_0'$:

$$\alpha = 1 - \exp[-k^{m+1}(t + t_0')^{m+1}] \quad . \tag{16}$$

where $t_0'$ is the time necessary for the isothermal regime to reach its initial transformed fraction, i.e. the transformed fraction after the constant heating regime:

$$t_0' = \left(\frac{\alpha_{ex,0}}{k}\right)^{1/m+1} - t_0 \tag{17}$$

where $\alpha_{ex,0}$ is the corresponding initial extended transformed fraction. This analysis is the basis of Krüger's numerical approach.[21] Nevertheless, this approach is generally wrong since the state of the system depends on the thermal history,[20,28] i.e. a given value of $\alpha$ will correspond to a different state and consequently it will evolve at a different rate. From Fig. 1 one can observe that the numerical solution and the approximate analytical solution, Eq. (A.2) appendix A, are practically identical (discrepancies are less than $10^{-4}$) while Eq. (17) gives, as expected, an incorrect prediction. The initial transformed fraction is 0.0036 whereas the error of the prediction according Eq. (17) is as high a 0.05. This result indicates that the deviations are significant even when the initial transformed fraction is below the detection level of most experimental setups. Indeed, the number of nuclei previously formed has a minor effect on the transformed fraction but has a pronounced effect on the subsequent crystallization evolution.

It is worth mentioning that our numerical approach keeps all the information related to the system such as temperature, transformed fraction, number of nuclei and size of the extended grains. Consequently, it can be applied to any arbitrary thermal history.

## IV. ALGORITHM FOR EVALUATION OF THE MORPHOLOGY

In this section, we present an algorithm which evaluates the grain morphology from the calculation of the extended transformed volume of each nucleus. The real volume $V$ is divided into an $m$-dimensional lattice formed by $m$-cubic cells of side $\Delta x$. Each cell is identified by a set of $m$ integer coordinates; the actual position of the cell is obtained by multiplying the integer coordinates by $\Delta x$. Initially a value of 0 is given to each cell; 0 stands for an untransformed cell. The evolution of the grain extended volume is performed according to the method described in Section II. When a new nucleus is created, a cell is assigned to this nucleus. The integer coordinates of this cell are chosen randomly. Each nucleus is identified by an integer number $i$. The number $i$ is then assigned to the corresponding cell provided that the cell does not belong to another grain. If the cell already belongs to another grain, then we are dealing with a "phantom nucleus". The concept of phantom nuclei was introduced by Avrami. The phantom nuclei do not appear in the lattice and have no effect on the calculation of the grain morphology, so they are discarded. However, as pointed out by Sessa et al.,[12] they must be included in the calculation of the total extended transformed fraction performed in Section II.

The algorithm used for the grain morphology evolution will be explained with the help of Fig. 2 where black cells correspond to the nuclei, the circumferences indicate the size of the extended grains, the grey regions are the cells associated to a particular grain and white cells represent the untransformed volume. The extended radius of the circumferences, $r_i$, is calculated from Eqs. (7) and (11). When the extended radius increment approaches $\Delta x$ all untransformed cells are checked to verify whether they have been incorporated to a neighboring grain or not. To save computing time, only the grains that have at least one transformed cell "in the vicinity" of the center of the untransformed cell are checked. The vicinity analyzed is determined by $\Delta x$. If there is a grain, $i$, such that the distance from the nucleus (center of the black cell) to the center of the untransformed cell is smaller or equal to $r_i$, then the number $i$ is assigned to the cell. When the cell is in the range of more than one grain, then the grain which reaches the cell first is assigned to the cell. Otherwise the cell remains untransformed. For example, the dashed circumferences represent the extended volume grown during the next step. The cell $a$ remains untransformed since there are no grains in the vicinity. Cells $b$ and $c$ only have a grain in the vicinity, which is the dark grey. Cell *b* will turn dark grey since its distance to the nucleus is smaller than the circumference radius, and conversely, cell *c* will remain untransformed. Finally two grains are located in the vicinity of cell *d* and the distance to both nuclei is smaller than the circumference radius, so cell *d* belongs to the grain that reaches it first.

Once the grain morphology has been built, the grain size distribution can be calculated directly from the radius of a grain. For 3D grain growth:

$$\widetilde{r}_i \equiv \sqrt[3]{\frac{3v_i}{4\pi}} \qquad (18)$$

where $v_i$ is the actual grain volume. And the transformed fraction is:

$$\alpha(t) = \frac{1}{V} \sum_{i=1}^{N(t)} \frac{4}{3} \pi \widetilde{r}_i^{\,3}(t) \qquad (19)$$

Besides, nucleation and growth calculations are separate from the grain morphology evaluation. Indeed, $\Delta x$ is usually several orders of magnitude larger than $\Delta r$ and one single microstructure evaluation involves a large number of time steps. Therefore, our algorithm can easily deal with time dependent nucleation and growth rates. Actually, the computer time required to evaluate nucleation and growth (Section II) is practically negligible when compared to the microstructure calculation. Hence, handling complex time dependences for nucleation and growth rates does not represent a significant amount of the computing time.

## V. ACCURACY OF THE ALGORITHM

In the Monte Carlo and cellular automata calculation methods, the principal source of error is the space discretization. The growth at each step is associated to one cell. To minimize this error, the linear growth is adjusted to be an integer fraction of $\Delta x$. The error can be avoided also by using particular growth modes.[29] However, in other cases it is unavoidable that, at each step, some cells are incorrectly assigned. The result is an accumulative error that progressively reduces the accuracy. Consequently, $\Delta x$ must be as small as possible. $V$ should be high enough to reduce the boundary effects and to allow a high population of nuclei to minimize statistical errors. Taking into account the computer memory limitations, the latter restriction is especially dramatic in 3D simulations where the number of cells is $V/\Delta x^3$. Usually, satisfactory accuracy is reached by performing the calculation several times.[13] Indeed, averaging for *n* calculations is equivalent to involve *n* times more nuclei, thus the statistical error is reduced.

In our algorithm, since the grain growth is driven by the evolution of its extended volume, the accumulative error associated to $\Delta x$ is suppressed. A grain boundary is not allowed to grow unless its distance to the nuclei is less than or equal to $r_i$. Moreover, since $r_i$ is calculated accurately (according to Section II), the contribution of $\Delta x$ to the error is drastically reduced. Then, when compared to Monte Carlo or cellular automata calculations, our algorithm gives an accurate grain-size distribution for a relative coarse space discretization. Consequently, for a given number of cells, the number of nuclei used in the calculation is greater. Therefore, accurate results are obtained in just one run with minimal computer requirements.

To check the efficiency of our algorithm, we have simulated the 3D crystallization of amorphous silicon under isothermal conditions, i.e. constant nucleation and growth rates. The calculation has been performed in a $760^3$ lattice (432 million cells). The results are summarized in Figs. 3, 4 and 5. It is worth mentioning that both, the transformed fraction evolution and the grain-size distribution represented in Figs. 3 and 4, are independent of the particular values *I* and *G*, i.e. one can obtain the solution of any particular system simply by multiplying space and time by the scaling factors (Appendix B):

$$\tau = \left(I\,G^3\right)^{-1/4} \quad \text{and} \quad \lambda = \left(\frac{G}{I}\right)^{1/4} \tag{20}$$

Thus our results (Figs. 3, 4 and 5) can be directly compared to those of Crespo et al.[14] (Figs. 1, 2 and 3). As far as we now, the simulation of Crespo et al. is one of the most efficient Monte Carlo algorithms. Indeed, in a remarkable analysis of the effect of the spatial resolution, Pusztai et al.[13] showed that at least $2^{12}$ cells where necessary to obtain a precise result. Although Crespo et al. used a significantly lesser number of cells (16.8 million cells); they obtained a notably accurate evolution of the transformed fraction (Fig. 1 in ref. 14).

The number of cells in our simulation is larger than that of Crespo et al. However, they have performed the average for 32 simulations while we have done only one calculation, i.e. their calculation is equivalent to a single calculation with the same discretization but a number of cells of 32 times greater (537 million cells). Despite the fact that their total number of cells is slightly larger, our calculation is significantly more accurate. For instance, when comparing the evolution of the transformed fraction (Fig. 3); the largest deviation from the theoretical value in our case is $3 \cdot 10^{-4}$, while Crespo et al. reported an accuracy of $10^{-3}$. For the final grain radius distribution, Fig. 4.c, our simulation exhibits an error clearly smaller than the one obtained by Crespo et al. (Fig. 2.f in ref. 14). Moreover, from Fig. 4 one can observe that the accuracy of the grain radius distribution improves as the transformation proceeds. The reason is that, as the transformation proceeds, the number of nuclei increases and consequently the error diminishes. In opposition, in the case of Crespo et al. (Fig. 2 in ref. 14), the accuracy diminishes as the transformation proceeds. The reason is the accumulative error related to the space discretization in the Monte Carlo method. Thus, one can conclude that the accumulative error due to the space discretization has been eliminated. Concerning the computer requirements, the program was run on a standard personal computer and lasted 36 minutes and the amount of memory required was 2 Gbytes.

Although our algorithm is able to work with larger values of $\Delta x$, its accuracy is still limited by the space discretization. Indeed, the smaller the grain is, the greater the error introduced by $\Delta x$. To minimize this effect, we have introduced the following condition: if the grain radius of Eq. (18) is larger than the extended radius $r_i$, then we take, $\tilde{r}_i = r_i$, otherwise the initial value of Eq. (18) is taken. This condition significantly reduces the error introduced by $\Delta x$ at the first stages of nuclei growth but has no effect when the grain impingement takes place. When the grains impinge, the precise calculation of $r_i$ allows us to accurately establish which grain reaches the center of a particular cell first. However, the center of a cell belonging to a grain does not necessarily apply to the rest of the cell. Thus the error in the calculation of the grain radius is about 50% of $\Delta x$. Therefore, $\Delta x$ must be, at least, one order of magnitude smaller than $<\tilde{r}>$. On the other hand, the larger $\Delta x$ is, the larger the number of nuclei and the smaller the statistical error is. Actually, the analysis of the accuracy of the algorithm with respect to $\Delta x$ confirmed that the optimum value of $\Delta x$ is approximately $0.1<\tilde{r}>$.

## VI. MICROSTRUCTURE AND GRAIN-SIZE DISTRIBUTION UNDER ISOTHERMAL CONDITIONS

In this Section we will analyze the different microstructures that emerge depending on the nucleation mechanism under isothermal conditions. We will focus our attention on the case of amorphous silicon. In this case, nucleation is continuous. In addition, the nucleation mechanism can be modified by introducing pre-existing nuclei, e.g. by ion implantation prior to isothermal annealing[30] or by pre-annealing the sample.[31] Furthermore, the simultaneous nucleation by both mechanisms has also been observed in the crystallization of metallic glasses.[32] Thus, we will analyze the site saturation nucleation case (crystallization of preexisting nuclei) alone and mixed with the continuous nucleation. As will be stated, the results obtained here can be extrapolated to any system featuring these nucleation mechanisms. In particular, the results obtained for continuous nucleation and pre-existing nuclei are universal.

*1 Continuous nucleation*

Under isothermal conditions, both *I* and *G* are constant. Hence, the system has an exact dimensional scaling, Eq. (20) (see Appendix B), so the behavior is universal.[33] In Figs. 4 and 5 we show the final grain radius distribution and an image of a central section respectively. To assess the accuracy of the grain size distribution, we have calculated the final average grain size and compared it to the analytical value. For the particular case of 3D growth, the average grain size is defined as:

$$<\tilde{r}> \equiv \sqrt[3]{\frac{1}{N}\sum_{i=1}^{N}\tilde{r}_i^3} \qquad (21)$$

and can be calculated through the simple relationship (Appendix C):

$$<\tilde{r}> = \sqrt[3]{\frac{3V}{4\pi N}} \qquad (22)$$

We have obtained a value of $<\tilde{r}> = 0.6436\lambda$ which matches with extraordinary accuracy the analytical exact value of $0.6435\lambda$ (Appendix C). This means that the distribution obtained for the final transformed state is very accurate. In fact, it is by far the most accurate so far published. The same test has been done for a 2D growth and gives an average grain size of $0.6018\lambda$ while the exact value is $0.6016\lambda$. The corresponding grain radius distribution for 2D growth is shown in Fig. 8.

Concerning the grain radius dispersion, defined by the standard deviation of the distribution:

$$\sigma_r \equiv \sqrt{\frac{1}{N}\sum_i (\tilde{r}_i - \bar{r})^2} \quad \text{and} \quad \bar{r} \equiv \frac{1}{N}\sum_i \tilde{r}_i \qquad (23)$$

it is quite large. For the completely transformed state ($\alpha = 1$) it is $0.485\,\bar{r}$ (3D) and $0.490\,\bar{r}$ (2D). It is worth noting that for the 2D case, the transformed fraction calculated from the grain distribution coincides with the exact solution within an accuracy of $8\cdot 10^{-4}$, while in ref. 13 for a significantly large number of cells, $2^{12}$, the maximum error is 0.02.

### 2 Pre-existing nuclei

In this case, with the natural time and space scaling,[27]

$$\tau' = (n_0\,G^3)^{-1/3} \quad \text{and} \quad \lambda' = \left(\frac{1}{n_0}\right)^{1/3} \qquad (24)$$

the dimensionless solution is universal, as well. The parameter $n_0$ is the pre-existing nuclei density. Fig. 7 shows the corresponding final grain radius distribution. Its average grain size is obviously $<\tilde{r}> = \sqrt[3]{3/4\pi}\,\lambda' = 0.62035\lambda'$. From the resulting microstructure we have calculated its standard deviation: $0.145\bar{r}$, where $\bar{r} = 0.608\lambda'$. Therefore the size distribution is significantly narrower than in the preceding case because all nuclei appear simultaneously.

This distribution fits a Gaussian distribution with remarkable accuracy (the square correlation coefficient is 0.9998):

$$y = \frac{1}{\sqrt{2\pi}\sigma} e^{-\frac{(x-\mu)^2}{2\sigma^2}} \qquad (25)$$

where the fitted parameters are $\mu = 0.609\lambda'$ and $\sigma = 0.0893\lambda'$. Note that the fitting parameters are in good agreement with $\bar{r} = 0.608\lambda'$ and the standard deviation $0.0883\lambda'$, respectively.

The final grain radius distribution for 2D growth is plotted in Fig. 8. Here again we fit the calculated distribution to a Gaussian distribution, though the fit is not as fine (the square correlation coefficient is 0.997). The fitted parameters are $\mu = 0.535$ and $\sigma = 0.148$ while $\bar{r} = 0.545\lambda'$ and the standard deviation is $0.146\lambda'$.

### 3 Continuous nucleation combined with pre-existing nuclei

As outlined in appendix C, there is not a universal solution in this case. Instead, the grain size distribution depends on the relative contribution of both nucleation mechanisms. Indeed, in appendix C it is shown that the final mean grain size depends only on two parameters: $n_0' \equiv n_0/n$ (where $n$ is the density of grains that would result without preexisting nuclei) and the space scaling factor $\lambda$. In Fig. 9 the final average grain size calculated from the grain size distribution and from the theoretical prediction (Eqs. (C.4) and (C.14)) are plotted against $n_0'$. Here again the agreement is excellent. The grain size diminishes monotonically since the number of nuclei increases with $n_0'$. Moreover, when $n_0'$ increases, the role of pre-existing nuclei is more relevant and the average grain size approaches the exact solution of nucleation driven only by pre-existing nuclei (dashed curve on Fig. 9).

Concerning the distribution, from Fig. 10 one can distinguish a narrow distribution of large grains due to the pre-existing nuclei and a continuous band of smaller grains related to the continuous nucleation. Here again, when $n_0'$ is high the effect of pre-existing nucleation is more noteworthy. In fact (see Fig. 9), the distribution width decreases monotonically as $n_0'$ increases.

Now, we can address the following question: How can one control the final microstructure under isothermal conditions? Concerning the average grain size, for continuous nucleation, it is proportional to a factor that depends on both nucleation and growth rates. Thus, once we know their temperature dependence, we can easily control the final grain size by choosing the appropriate temperature. However, since the solution is universal, we cannot control the grain size distribution (shape and width). On the other hand, the distribution is significantly narrower when only pre-existing nuclei are present. Here again the mean grain size can be easily controlled with temperature, but the distribution is also universal. Thus, the distribution can be modified by mixing both nucleation mechanisms. However, in this case, the resulting distribution is very similar to a superposition of the distributions corresponding to both nucleation mechanisms acting independently.

Although the results have not been detailed for all the simulations reported, the agreement between the transformed fraction calculated numerically or analytically from Eq. (4) and from the constructed microstructure is very good (the error is smaller than 0.003). These results, in addition to those of Sessa et al.,[12] Crespo et al.[14] and Pusztai et al.,[13] represent a direct confirmation of the Avrami theory (in particular of Eqs. (1) and (4)) for a number of particular cases.

## VII. CONCLUSIONS

To sum up, we have introduced a new numerical method for solving the kinetic equations of KJMA theory. Our method, shares the main Avrami assumptions: random nucleation and isotropic growth. The numerical model takes into account all the parameters that define the system state and no restrictions are imposed either to the thermal history or to the growth and nucleation rate dependencies on temperature. A comparison between the numerical results and the analytical solutions speaks for the high accuracy of the method (relative error smaller than $10^{-6}$).

In addition, we have introduced an algorithm to calculate the grain morphology from the extended microstructure. Since the extended microstructure is calculated according the previous numerical model, this algorithm keeps its flexibility and can be easily adapted to any conditions. In contrast, the calculation of the transformed fraction is not based on KJMA theory. Thus, its applicability is not restricted by the assumptions of KJMA theory. This fact means that our algorithm can be used to test the correctness and range of applicability of KJMA's theory. Compared to existing numerical methods, our algorithm features high accuracy in relatively short computational times and low computer memory requirements.

The algorithm has been used to obtain the universal grain size distribution for constant growth rate and constant continuous nucleation as well as for pre-existing nuclei. As far as we know, both results are the most accurate ever published. In addition, the case of continuous nucleation combined with pre-existing nuclei has been analyzed. In this case, a universal solution does not exist. However, the grain size distribution depends on the relative contribution of both mechanisms (apart from a size scaling factor).

In all cases, the agreement between numerical results and theoretical predictions is excellent. Table II summarizes the calculated average grain size and the standard deviation of the grain radius distribution.

## ACKNOWLEDGMENTS

This work has been supported by the Spanish *Programa Nacional de Materiales* under contract number MAT2006-11144.

**APPENDIX A. ANALYTICAL SOLUTION FOR AN ISOTHERMAL REGIME SUBSEQUENT TO CONSTANT A HEATING STEP.**

During the isothermal regime, the extended transformed fraction has two contributions, one from the nuclei created during the heating step and a second one from the nuclei created during the isothermal step:

$$\alpha_{ex}(t) = \sigma I \int_0^{t_0} r^m(t,\tau) d\tau + \sigma I \int_{t_0}^{t} G^m (t-\tau)^m d\tau \tag{A.1}$$

where $t_0$ is the time where the isothermal regime starts and $r(t,\tau)$ is the radius at a time $t$ of a nucleus created at a time $\tau$. Actually, $r(t,\tau)$ is the result of the sum of the growth during the heating rate and the growth during the isothermal regime. The first term of Eq. (A.1) is calculated using the same approximation of ref. 19 and its solution is the first term of Eq. (A.2). The second term of Eq. (A.2) is the exact solution of the second contribution and is the well known solution of the isothermal case:[7]

$$\alpha_{ex}(t) = k_0^{m+1} \left(\frac{E}{\beta k_B}\right)^{m+1} \left\{ \sum_{j=1}^{m+1} \left[ \binom{m+1}{j} \frac{j! E^j}{\prod_{i=0}^{j-1}(E_N + iE_G)} \left(\frac{E_N + (j-1)E_G}{jE}\right)^{2j} \left[ p\left(\frac{E_N + (j-1)E_G}{jk_B T}\right) \right]^j \right. \right.$$
$$\left. \left. \times \left(\frac{\beta k_B}{E} \exp\left[-\frac{E_G}{k_B T}\right](t-t_0)\right)^{m+1-j} \right] \right\} + k^{m+1}(t-t_0)^{m+1} \tag{A.2}$$

where $\beta$ is the heating rate and $p(x) \equiv \int_x^{\infty} \frac{\exp(-u)}{u^2} du$.

When $E_N = E_G$, Eq. (A.2) reduces to:

$$\alpha_{ex}(t) = k^{m+1}(t + t'_0)^{m+1} \tag{A.3}$$

Substituting Eq. (A.3) into Eq. (4) one obtains Eq. (16). Indeed, when nucleation and growth have the same evolution, the system state only depends on the transformed fraction.[19] Thus when $E_N = E_G$ the system evolution does not depend on the particular thermal history and Eq. (16) is valid.

**APPENDIX B. DIMENSIONAL SCALING FOR CONSTANT NUCLEATION AND GROWTH RATES.**

Let's assume that two systems, characterized by their particular values of the growth and nucleation rates ($G_1$, $I_1$, $G_2$, and $I_2$), differ only by a scale factor at given times, $t_1$ and $t_2$, respectively (see Fig. 11). Their microstructure will maintain this scale relation provided that at $t_1 + dt_1$ and $t_2 + dt_2$ the following conditions are fulfilled: *(i)* the interface advances proportionally to the system length, $L_i$,

$$\frac{dx_1}{L_1} = \frac{dx_2}{L_2} \quad (B.1)$$

and *(ii)* the number of new formed nuclei in the untransformed volume is the same,

$$d\Delta N_1 = d\Delta N_2 \quad (B.2)$$

bearing in mind that,

$$dx_i = G_i dt_i \quad \text{and} \quad d\Delta N_i = L_i^3(1-\alpha)I_i dt_i \quad (B.3)$$

where $\alpha$ is the transformed fraction (the same for both systems at $t_1$ and $t_2$), Eqs. (B.1) and (B.2) become:

$$\frac{G_1 dt_1}{L_1} = \frac{G_1 dt_2}{L_2} \quad \text{and} \quad L_1^3 I_1 dt_1 = L_2^3 I_2 dt_2 \quad (B.4)$$

from which,

$$\frac{dt_1}{\tau_1} = \frac{dt_2}{\tau_2} \quad \text{and} \quad \frac{L_1}{\lambda_1} = \frac{L_2}{\lambda_2} \quad (B.5)$$

where $\tau_i \equiv (I_i G_i^3)^{-1/4}$ and $\lambda_i \equiv \left(\frac{G_i}{I_i}\right)^{1/4}$ are respectively the time and space scale factor. Thus, the behavior of both systems is the same when scaled by $\tau$ and $\lambda$.

Finally, note that under the previous scaling, the dimensionless growth and nucleation rates are equal to 1. Indeed, the fact that dimensionless parameters do not depend on the particular values of *I* and *G* proves that the dimensionless system is universal, i.e. the evolution of any particular system can be obtained from the dimensionless system simply by multiplying the dimensionless time and space by $\tau$ and $\lambda$ respectively.

**APPENDIX C. AVERAGE GRAIN SIZE OF THE FULLY CRYSTALLIZED MATERIAL.**

For the sake of simplicity, the calculations are done for 3D growth only. The mean grain size is defined as:

$$<\tilde{r}> \equiv \sqrt[3]{\frac{1}{N}\sum_{i=1}^{N}\tilde{r}_i^3} \qquad (C.1)$$

where $N$ is the total number of grains. Likewise, when the transformation is over, the total volume is the sum of the volume of the individual grains, $v_i$:

$$V = \sum_{i=1}^{N} v_i \qquad (C.2)$$

and taking into account the definition of the grain radius (Eq. (18)),

$$V = \frac{4}{3}\pi \sum_{i=1}^{N} \tilde{r}_i^3 \qquad (C.2)$$

combining Eqs. (C.1) and (C.3), one gets

$$N\frac{4}{3}\pi <\tilde{r}>^3 = \frac{4}{3}\pi \sum_i N_i \tilde{r}_i^3 = V \qquad (C.3)$$

whence,

$$<\tilde{r}> = \sqrt[3]{\frac{3}{4\pi}\frac{V}{N}} \qquad (C.4)$$

Let's first analyze the case of pre-existing nuclei. The total number of nuclei is constant $N = n_0 \cdot V$, thus

$$<\tilde{r}>_{3D} = \sqrt[3]{\frac{3}{4\pi}\frac{1}{n_0}} = 0.62035 \cdot (n_0)^{-1/3} \qquad (C.5)$$

where $(n_0)^{-1/3}$ is the corresponding dimensional space scaling.

For 2D growth the result would be:

$$<\tilde{r}>_{2D} = \sqrt[2]{\frac{1}{\pi}\frac{1}{n_{0,2D}}} \approx 0.56420 \cdot (n_0)^{-1/2} \qquad (C.6)$$

For the case of continuous nucleation, the total number of nuclei is equal to

$$N(t) = I \cdot V \int_0^t 1 - \alpha(u) du \qquad (C.7)$$

In other words, nuclei become real grains only when they appear in the untransformed volume. Then, for isothermal and continuous nucleation, Eq. (C.7) reduces to:

$$N(t) = I \cdot V \int_0^t \exp(-k^4 u^4) du = I \cdot V \int_0^t \exp\left(-\frac{\pi}{3}\frac{u^4}{\tau^4}\right) du \quad (C.8)$$

and the final number of nuclei is

$$N = I \cdot V \int_0^\infty \exp\left(-\frac{\pi}{3}\frac{u^4}{\tau^4}\right) du = V\left(\frac{3}{\pi}\right)^{1/4} \lambda^{-3} \Gamma\left(\frac{5}{4}\right) \quad (C.9)$$

thus

$$<\tilde{r}>_{3D} \approx 0.6435 \lambda \quad (C.10)$$

Since both dimensionless systems are universal, the mean grain radius is proportional to the dimensional space scaling factor as expected. For 2D growth the result would be:

$$N = I \cdot V \int_0^\infty \exp\left(-\frac{\pi}{3}\frac{u^3}{\tau^3}\right) du = V\left(\frac{3}{\pi}\right)^{1/3} \lambda^{-2} \Gamma\left(\frac{4}{3}\right) \text{ and } <\tilde{r}>_{2D} \approx 0.6016 \lambda \quad (C.11)$$

Finally, we should point out that there is not an analytical solution for the case where both nucleation mechanisms are mixed. The final number of grains is:

$$N = IV \int_0^\infty \exp\left[-\frac{4\pi}{3}\left(\frac{t^4}{4\tau^4} + n_0 G^3 t^3\right)\right] dt + n_0 V \quad (C.12)$$

If we define $n_0'$ as the ratio between the density of preexisting nuclei, $n_0$, and the final density of grains of the system when transformed without preexisting nuclei, $n$:

$$n_0' \equiv n_0 / n \quad (C.13)$$

then, substitution of (C.9) in (C.12) gives:

$$N = \frac{V}{\lambda^3}\left(\int_0^\infty \exp\left[-\frac{4\pi}{3}\left(\frac{u^4}{4} + \frac{n_0' u^3}{\kappa}\right)\right] du + \frac{n_0'}{\kappa}\right) \quad (C.14)$$

where $\kappa \equiv \left(\left(\frac{3}{\pi}\right)^{1/4} \Gamma\left(\frac{5}{4}\right)\right)^{-1} \approx 1.1161$. Hence, the mean radius depending on two parameters, $\lambda$ and $n_0'$. $n_0'$, accounts for the relation between both nucleation mechanisms. The larger $n_0'$ is, the more important the contribution of the pre-existing nuclei.

**Table I**. Experimental nucleation and growth rates of amorphous silicon[1].

| | | |
|---|---|---|
| Nucleation | Activation energy | 5.3 eV |
| | Preexponential term | 1.7 $10^{44}$ $s^{-1}$ $m^{-3}$ |
| Growth | Activation energy | 3.1 eV |
| | Preexponential term | 2.1 $10^{7}$ $s^{-1}$ m |

**Table II**. Analytical and calculated average grain size and grain radius distribution dispersion.

| | | Analytical 3D | Calculated 3D | Analytical 2D | Calculated 2D |
|---|---|---|---|---|---|
| Dimensionless $<\tilde{r}>$ | Continuous | 0.6435 | 0.6436 | 0.6016 | 0.6018 |
| Standard deviation | nucleation | - | 0.485 $\bar{r}$ | - | 0.490 $\bar{r}$ |
| Dimensionless $<\tilde{r}>$ | Pre-existing | 0.62035 | 0.62033 | 0.56419 | 0.56416 |
| Standard deviation | nuclei | - | 0.145 $\bar{r}$ | - | 0.269 $\bar{r}$ |

**Figure captions**

Figure 1. 3D isothermal crystallization of amorphous silicon at 700ºC after continuous heating at 20 K/min: our numerical method (crosses), the rough approximation of Eq. (17) (dotted line) and the quasi-exact solution of Eq. (A.2) (solid line). The difference between Eq. (17) and our numerical solution is shown by the dashed line (multiplied by 10). $\Delta r = 5 \cdot 10^{-5}$ μm, $V = 8 \cdot 10^3$ μm³ and total number of nuclei is 182.231.

Figure 2. Schematic representation of the algorithm used to calculate the microstructure.

Figure 3. Transformed fraction versus time for 3D growth of amorphous silicon at constant temperature, $T$=680ºC, and continuous nucleation. The solid line represents the exact solution, Eq. (14). Crosses correspond to the transformed fraction calculated from the microstructure. The time is normalized according to Eq. (19). Simulation parameters: $\Delta r = 5 \cdot 10^{-5}$ μm, $V = 12167$ μm³, number of cells $760^3$, total number of nuclei 188144 (89745 phantom) and $\Delta x = 0.03$ μm ($<\tilde{r}> = 0.3089 \mu m$).

Figure 4. Grain radius distribution for 3D growth of amorphous silicon at three different transformed fractions at constant temperature, $T$=680ºC, and continuous nucleation. Simulation parameters are given in the caption of Fig. 3. With the normalized radius (Eq. (19)) these distributions are universal.

Figure 5. Simulated microstructure cross section corresponding to the crystallization of amorphous silicon at three different transformed fractions at constant temperature, $T$= 680ºC, and continuous nucleation. Simulation parameters are given in the caption of Fig. 3. With proper time and space normalization these grain morphologies are universal.

Figure 6. Final grain radius distribution for 2D growth of amorphous silicon at constant temperature, $T$=680ºC, and continuous nucleation. Simulation parameters: $\Delta r = 10^{-5}$ μm, $A = 40000$ μm², number of cells $20000^2$, total number of nuclei 579460 (330840 phantom) and $\Delta x = 0.010$ μm ($<\tilde{r}> = 0.2263 \mu m$). With the normalized radius (Eq. (24)) this distribution is universal.

Figure 7. Final grain radius distribution for 3D crystallization of amorphous silicon due to preexisting nuclei ($n_0 = 8.69\,\mu m^{-3}$) at $T$=680ºC. The solid line is the corresponding Gaussian fit. With the normalized radius, the distribution is universal. Simulation parameters: $\Delta r = 5 \cdot 10^{-5}\,\mu m$, $V = 10648\,\mu m^3$, number of cells $760^3$ and $\Delta x = 0.030\,\mu m$ ($<\tilde{r}> = 0.3017\,\mu m$).

Figure 8. Final grain radius distribution for 2D crystallization of amorphous silicon due to preexisting nuclei ($n_0 = 64\,\mu m^{-2}$) at $T$=680ºC. Simulation parameters: $\Delta r = 5 \cdot 10^{-6}\,\mu m$, $A = 7225\,\mu m^2$, number of cells $2000^2$ and $\Delta x = 0.0042\,\mu m$ ($<\tilde{r}> = 0.0705$).

Figure 9. Final average grain size at constant temperature, $T$=680ºC, for continuous nucleation combined with pre-existing nuclei. Squares are calculated from the grain size distribution while the solid line corresponds to the numerical solution of the theoretical prediction (Eqs. (C.4) and (C.14)). The dashed line is the theoretical solution when crystallization is due to pre-existing nuclei only (Eq. (C.5)) and the dotted line is the relative standard deviation. The curves are universal.

Figure 10. Final grain radius distribution at constant temperature, $T$=680ºC, for continuous nucleation combined with pre-existing nuclei. Grey bars $n_0' = 0.558$ and black bars $n_0' = 3.348$.

Figure 11. Schematic representation of two scalable microstructures. New nuclei are depicted with crosses.

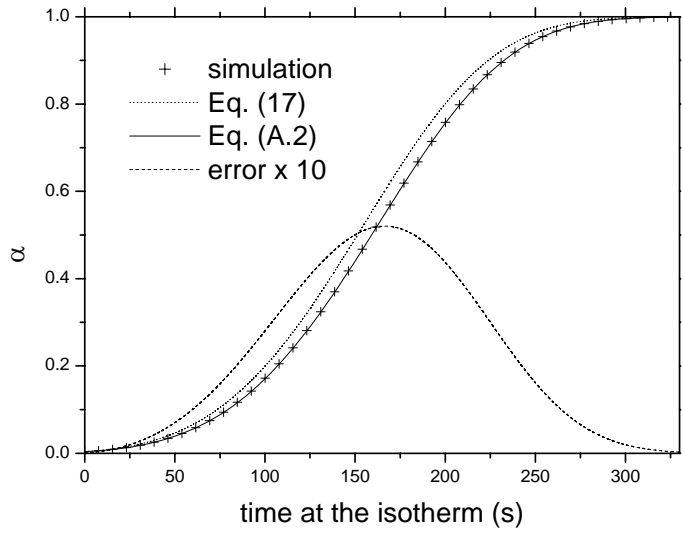

Figure 1

Figure 2

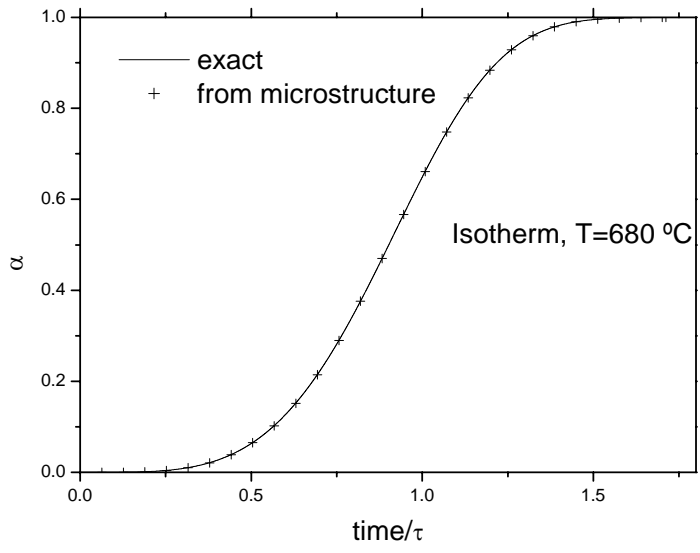

Figure 3

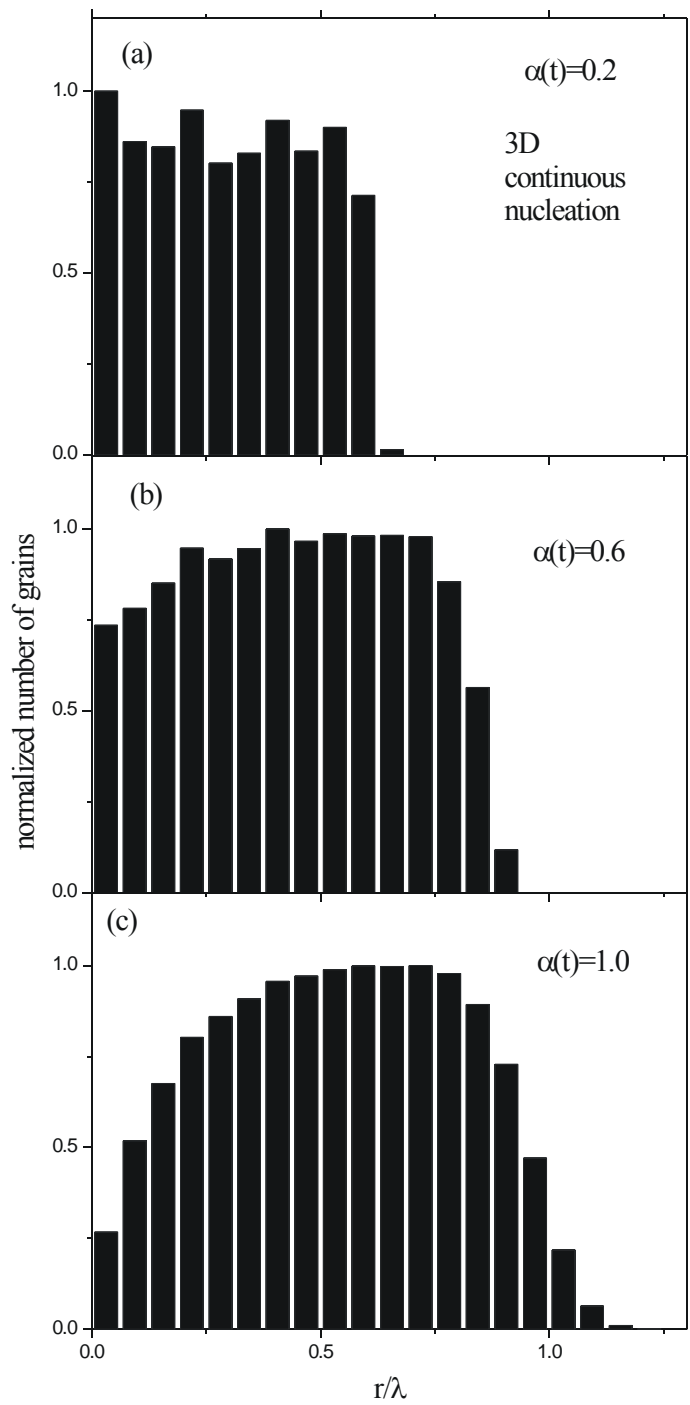

Figure 4

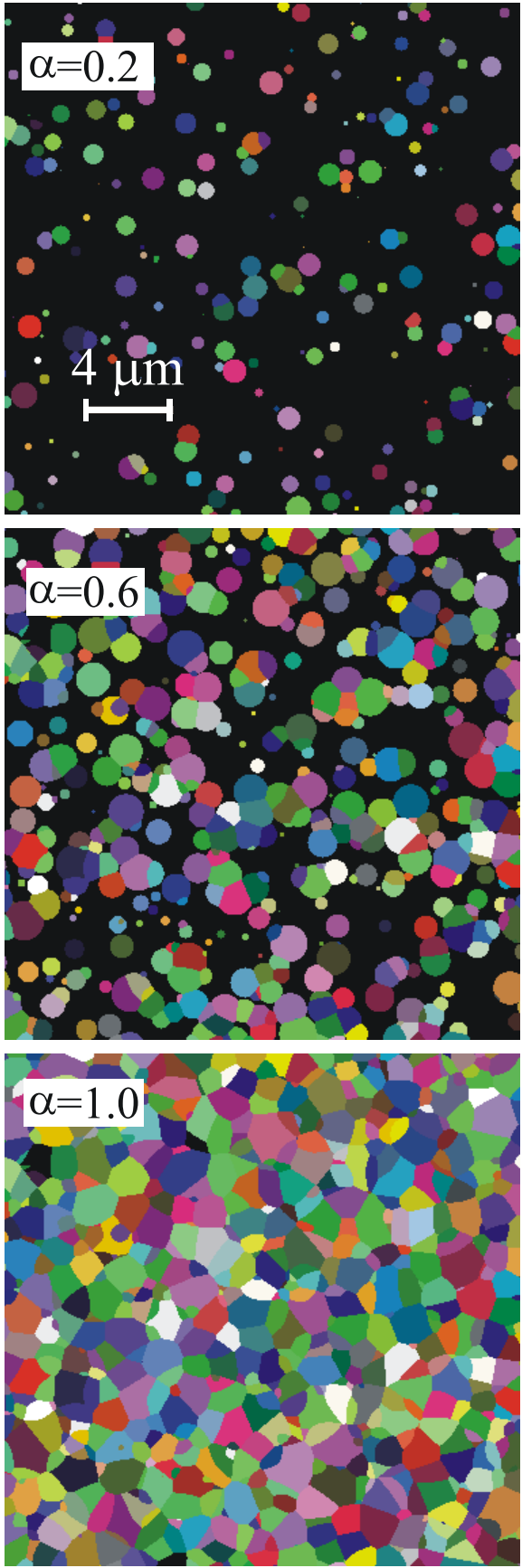

Figure 5

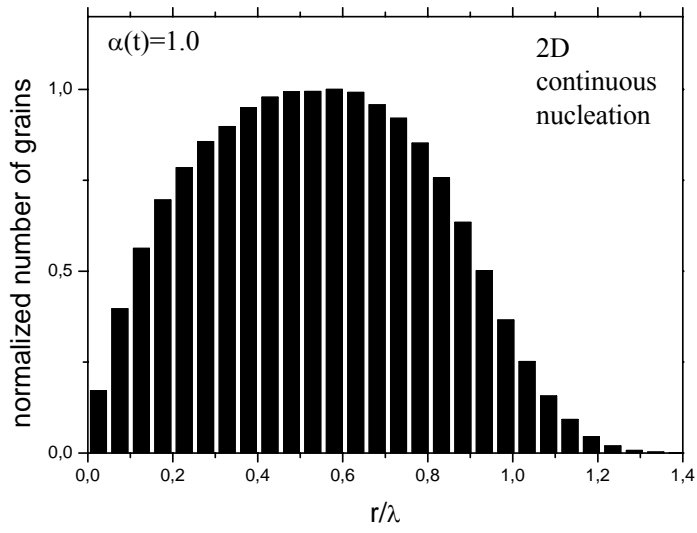

Figure 6

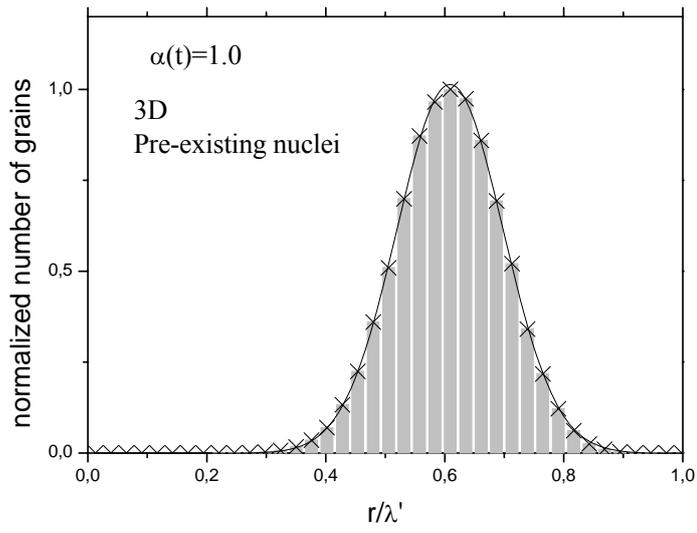

Figure 7

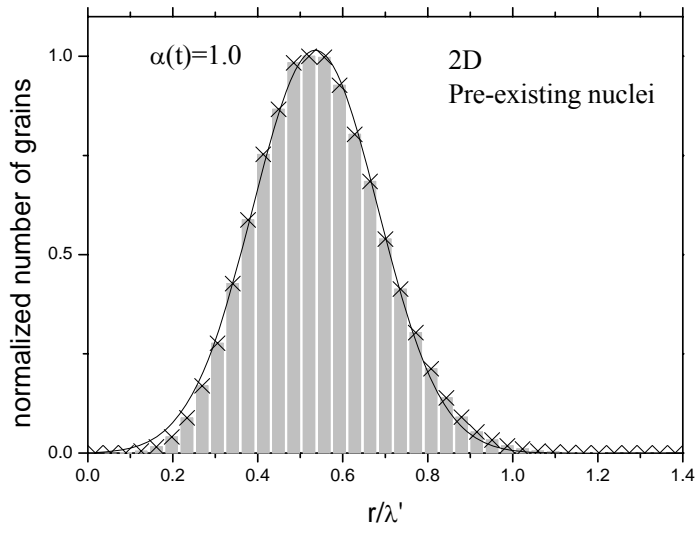

Figure 8

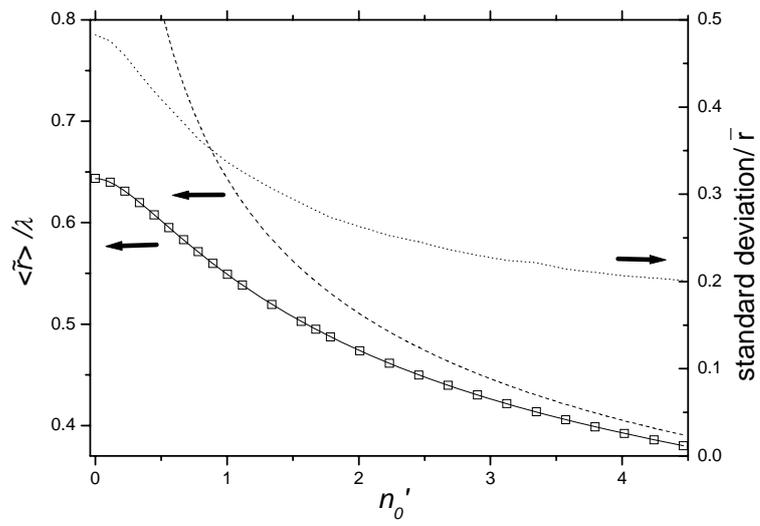

Figure 9

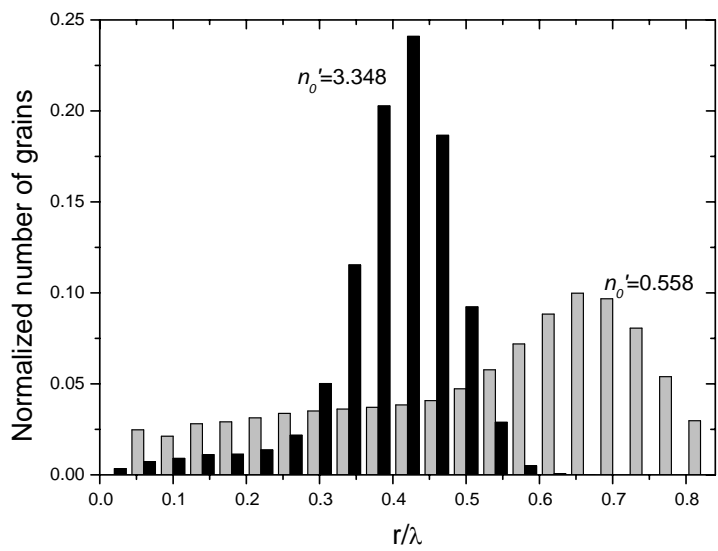

Figure 10

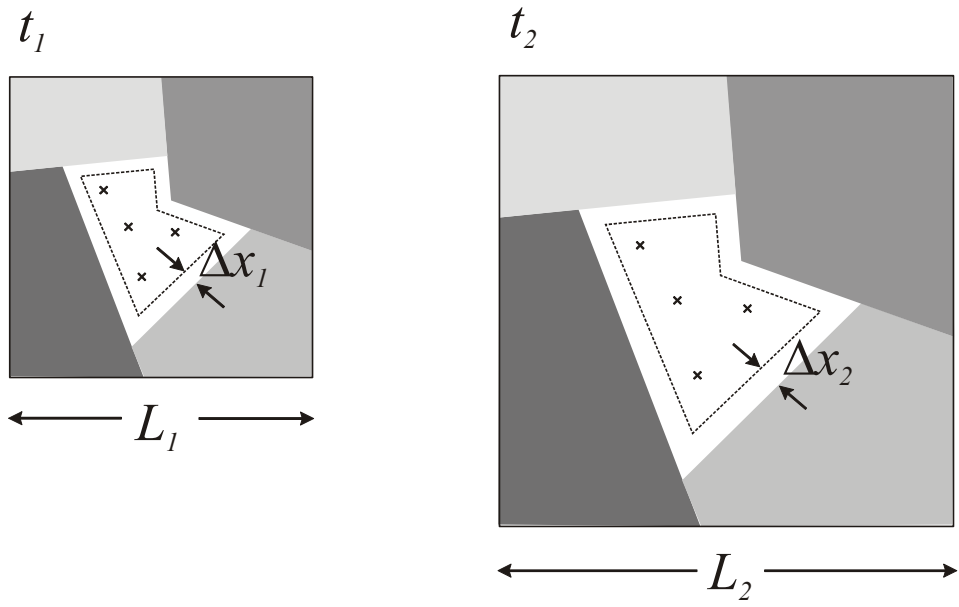

Figure 11